\documentclass{PoS}                              

\usepackage[latin1]{inputenc}                    
\usepackage{graphicx}                            
\usepackage{latexsym}                            
\usepackage{amsfonts}                            
\usepackage{amssymb}                             
\usepackage{amsmath}                             
\usepackage[mathscr]{eucal}                      
\usepackage{dcolumn}                             

\PoS{PoS(LAT2005)056}                            

\newcommand{\imag}{\mbox{i}}                     
\newcommand{\msbar}{$\overline{\mbox{\rm MS}}$}  
\graphicspath{{.}{Graphics/}{plots/}}            

\title{Hadron structure with light dynamical quarks}

\ShortTitle{Hadron structure with light dynamical quarks}

\author{Robert G.~Edwards\\Thomas Jefferson National Accelerator
  Facility, Newport News, VA 23606, USA}

\author{George T.~Fleming\footnotemark[1]\\Sloane Physics Laboratory,
  Yale University, New Haven, CT 06520, USA}

\author{Philipp H{\"a}gler\\Department of Physics and Astronomy, Vrije
  Universiteit, 1081 HV Amsterdam, NL}

\author{John W.~Negele\\Center for Theoretical Physics, Massachusetts
  Institute of Technology, Cambridge, MA 02139, USA}

\author{Kostas Orginos\\Center for Theoretical Physics, Massachusetts
  Institute of Technology, Cambridge, MA 02139, USA}

\author{Andrew Pochinsky\\Center for Theoretical Physics,
  Massachusetts Institute of Technology, Cambridge, MA 02139, USA}

\author{Dru B.~Renner\footnotemark[1]\\Department of Physics,
  University of Arizona, 1118 E 4th Street, Tucson, AZ 85721, USA}

\author{David G.~Richards\\Thomas Jefferson National Accelerator
  Facility, Newport News, VA 23606, USA}

\author{\speaker{Wolfram Schroers\footnotemark[1]}\\John von
  Neumann-Institut f{\"u}r Computing NIC/DESY, 15738 Zeuthen, Germany}

\abstract{Generalized parton distributions encompass a wealth of
  information concerning the three-dimensional quark and gluon
  structure of the nucleon, and thus provide an ideal focus for the
  study of hadron structure using lattice QCD. The special limits
  corresponding to form factors and parton distributions are well
  explored experimentally, providing clear tests of lattice
  calculations, and the lack of experimental data for more general
  cases provides opportunities for genuine predictions and for guiding
  experiment.  We present results from hybrid calculations with
  improved staggered (Asqtad) sea quarks and domain wall valence
  quarks at pion masses down to 350 MeV. \\ Preprint: MIT-CTP 3685,DESY 05-197}

\FullConference{XXIIIrd International Symposium on Lattice Field Theory\\
  25-30 July 2005\\Trinity College, Dublin, Ireland}

\begin{document}

\section{Introduction}
\label{sec:introduction} The discovery of generalized parton
distributions (GPDs) (see \cite{Muller:1998fv} for the pioneering
articles and \cite{Diehl:2003ny} for recent comprehensive reviews) for
the characterization of both exclusive and inclusive reactions has
stirred new interest in both experimentalists and theoreticians. For
the first time it was possible not only to describe seemingly
unrelated processes in terms of a single set of functions
characterizing a hadron, but also to incorporate information that was
not available before. The ability to compute the total quark
contribution to the nucleon spin \cite{Mathur:1999uf} plays on
important role in resolving the spin crisis.

These developments provide an ideal opportunity for contemporary
lattice QCD. While these calculations are still limited to quark
masses heavier than in nature they can provide invaluable qualitative
insight into the mechanisms of QCD. The pioneering lattice works on
this field were published simultaneously by the QCDSF
\cite{Gockeler:2003jf} and LHPC \cite{Hagler:2003jd} collaborations.
Already these early papers are playing an important role in
phenomenology~\cite{Diehl:2004cx}.

Later lattice investigations have unraveled important information on
the transverse structure of the nucleon \cite{Hagler:2003is},
polarized processes \cite{Schroers:2003mf} and the tensor structure of
GPDs \cite{Gockeler:2005aw}. The role of lattice QCD is of particular
importance in this young field since the GPDs are inherently more
complicated than form factors or parton distributions alone. Several
of these functions can only be determined from lattice calculations
and are not directly accessible to experiments.

This paper focuses on the application of hybrid lattice calculations
\cite{Renner:2004ck} to the lowest moments of the GPDs. It is
organized as follows.  After an introduction to the parameterization
of GPDs in section~\ref{sec:gpds}, we outline the current status of
our hybrid calculations in section~\ref{sec:hybrid-calculations}.  We
present the preliminary results from our calculations in section
\ref{sec:gpd-moments} which covers the electromagnetic form factors
and the nucleon energy-momentum tensor. In the forward limit the GPDs
reduce to the moments of forward parton-distributions for which a
plethora of detailed experimental data is available. For this reason,
we discuss this case in more detail in section~\ref{sec:pdf-moments}.
Finally, we present an outlook to the analysis of form factors at
large momentum transfer in section~\ref{sec:form-factors}. We close
with our plans for future calculations to complete our program in
section~\ref{sec:summary-outlook}.

\section{Generalized parton distributions}
\label{sec:gpds}
In the following, we focus on the case of a nucleon. Naturally, GPDs
can be defined for all other hadrons as well. For the case of the
pion, see \cite{Brommel:2005la} in these proceedings. Generalized
parton distributions are defined by the nucleon matrix element of an
operator that creates and annihilates quarks separated by a given
distance on the light cone:
\begin{equation}
  \label{eq:gpd-def}
  \bar{p}^+ \int\frac{dz^-}{2\pi} e^{\mbox{i}\bar{p}^+ z^-} \langle
  p'\vert \bar{\psi}(-z^-/2)\Gamma\psi(z^-/2)\vert p\rangle
\end{equation}
with $p$ being the momentum of the incoming, and $p'$ being the
momentum of the outgoing nucleon. The average nucleon momentum is
$\bar{p}=1/2(p'+p)$. These matrix elements can be parameterized by
functions depending on three kinematic variables, the average
longitudinal momentum fraction, $x$, the skewness, $\xi$, and the
total spacelike virtual momentum transfer, $Q^2\equiv t=(p'-p)^2$.
Additionally, the GPDs implicitly depend also on a renormalization
scale, $\mu^2$.  This scale fixes the scale of the struck quark when
using GPDs in a factorization scheme. Depending on the Dirac
structure, $\Gamma$, in Eq.~(\ref{eq:gpd-def}) there are in total
eight GPD functions. These functions are summarized in
table~\ref{tab:all-ferm-gpds}.

\begin{table}
\begin{tabular}{r|c|l}\hline
  ${\Gamma}$ & Function & Meaning \\ \hline
  $\gamma^\mu$      & $H(x,\xi,t), E(x,\xi,t)$ & Spin-independent \\
  $\gamma_5\gamma^\mu$ & $\tilde{H}(x,\xi,t), \tilde{E}(x,\xi,t)$ &
  Spin-dependent \\
  $\sigma^{\rho\mu}\gamma^\nu$ & $H_T(x,\xi,t)$, $E_T(x,\xi,t)$, &
  Transverse \\
  & $\tilde{H}_T(x,\xi,t)$, $\tilde{E}_T(x,\xi,t)$ & \\ \hline
\end{tabular}
\caption{List of fermion GPDs for the nucleon.}
\label{tab:all-ferm-gpds}
\end{table}

Since we cannot compute non-local matrix elements in the Euclidean
regime where lattice calculations are performed, we have to compute
moments with respect to the longitudinal momentum fraction, $x$. These
moments are obtained by performing a light-cone operator product
expansion.  The resulting moments will then be polynomials in
$(2\xi)^2$. The coefficient functions are the so-called generalized
form factors (GFFs). In this paper, we restrict ourselves to moments
as high as the energy-momentum tensor of QCD. The parameterization
takes on the following form:
\begin{eqnarray}
  \label{eq:mom-eq}
  \int dx\, H(x,\xi,t) = A_{10}(t) = F_1(t)\,, &&
  \int dx\, E(x,\xi,t) = B_{10}(t) = F_2(t)\,, \\
  \int dx\, x H(x,\xi,t) = A_{20}(t) - (2\xi)^2 C_2(t)\,, &&
  \int dx\, x E(x,\xi,t) = B_{20}(t) + (2\xi)^2 C_2(t)\,.
\end{eqnarray}
The zeroth moment corresponds to the regular electromagnetic form
factors, while the first moment corresponds to the energy-momentum
tensor.  Furthermore, we restrict ourselves to the flavor combination
up minus down since in this case the flavor-singlet parts involving
disconnected diagrams cancel due to isospin symmetry.

\section{Hybrid lattice calculations}
\label{sec:hybrid-calculations}
The basic shortcoming of most currently employed lattice fermion
actions is their enormous cost which makes calculations with
sufficiently light sea quarks prohibitively expensive. The principal
idea behind hybrid calculation consists of using different types of
fermion actions for the sea quarks --- the virtual quark and
anti-quark pairs created from loops from the gluon propagator --- and
the valence quarks which connect to the source and the sink.

In the present work, we use Asqtad fermions for the sea quarks
from configurations generated by the MILC collaboration. We then use
domain wall fermions with an exact lattice
chiral symmetry for the valence quarks.  The resulting lattice theory
should become a valid description of the full theory in the continuum
limit provided this limit exists.  These simulations extend the
calculations we have already reported on in \cite{Renner:2004ck}. The
current status of our lattice simulations is summarized in
table~\ref{tab:hybrid-stat}. All working points are at a constant
lattice spacing corresponding to $a^{-1}=1.588$ GeV.

\begin{table}
\begin{tabular}{c|c|c|c}\hline
$\mathrm{Volume}\ {\Omega}$ & Configs. & $\mathbf{(am_q)^{\mbox{\tiny Asqtad}}}$
& $\mathbf{m_{\mbox{\tiny PS}}}$ / MeV \\ \hline
$20^3\times 32$ & $425$ & $0.050$ & $790(2)$ \\
& $350$ & $0.040$ & $693(3)$ \\
& $564$ & $0.030$ & $594(2)$ \\
& $486$ & $0.020$ & $492(2)$ \\
& $655$ & $0.010$ & $354(2)$ \\
$28^3\times 32$ & $271$ & $0.010$ & $352(1)$ \\ \hline
\end{tabular}
\caption{Summary of our working points and statistics for our hybrid
  calculations.}
\label{tab:hybrid-stat}
\end{table}

\section{Moments of parton distributions}
\label{sec:pdf-moments}
The lattice calculations of low moments of the nucleon parton
distributions play multiple roles in the effort to understand the
non-perturbative structure of hadrons in QCD.  First, the lowest
moments of the nucleon's parton distributions are interesting
observables reflecting that the quark spin provides only a small
fraction of the nucleon's spin and that only a small portion of the
nucleon's momentum is carried by the quarks.  Quark orbital motion and
gluon contributions must account for the missing nucleon spin, and
additionally the gluons must provide the remaining momentum within the
nucleon.  Therefore reliable non-perturbative QCD calculations of the
moments of parton distributions are essential to the theoretical
effort to understand nucleon structure.

Additionally the lattice calculation of moments of nucleon parton
distributions will provide a benchmark test for the calculation of
other hadronic observables because the calculations of moments can be
compared directly with the corresponding experimental measurements of
the nucleon.  As lattice calculations are performed at light enough
quark masses that controlled quantitative comparison with experimental
results is achieved, then we will have confidence in the calculation
of other hadronic matrix elements which may be poorly determined by
experiment.  As an example, the moments of generalized parton
distributions or the large $Q^2$ limit of nucleon form factors
discussed in Sections~\ref{sec:gpd-moments} and \ref{sec:form-factors}
will require varying levels of difficulty to measurement
experimentally and the successful calculation of moments of ordinary
parton distributions will allow for genuine predictions from lattice
calculations for these and other observables.

\paragraph{Lattice details}
The moments of parton distributions are determined by calculating
matrix elements of the twist two operators in lattice QCD.  For
example the moments of the unpolarized and longitudinally polarized
distributions are given by
\begin{eqnarray*}
  \langle x^{n-1}\rangle_q p^{\{\mu_1}\cdots p^{\mu_n\}} & = &
  \frac{1}{2} \langle p,S | \overline{q}\imag D^{\{\mu_1}\cdots
  \imag D^{\mu_{n-1}}\gamma^{\mu_n\}}q | p,S \rangle\,, \\
  \langle x^{n-1}\rangle_{\Delta q} S^{\{\mu_1}\cdots p^{\mu_n\}} & =
  & \frac{2}{n+1} \langle p,S | \overline{q}\imag D^{\{\mu_1}\cdots
  \imag D^{\mu_{n-1}}\gamma^{\mu_n\}}\gamma^5 q | p,S \rangle\,, \\
\end{eqnarray*}
where the moments are defined as
\begin{eqnarray*}
  \langle x^{n-1}\rangle_q & = & \int_{-1}^{1} \!\!dx\,\, x^n q(x) =
  \int_0^1 \!\!dx\,\, x^n (q(x)-\overline{q}(x))\,, \\
  \langle x^{n-1}\rangle_{\Delta q} & = & \int_{-1}^{1} \!\!dx\,\, x^n
  \Delta q(x) = \int_0^1 \!\!dx\,\, x^n (\Delta
  q(x)+\Delta\overline{q}(x))\,.
\end{eqnarray*}
These moments are encoded in the generalized form factors discussed in
Section~\ref{sec:gpds} as
\begin{eqnarray*}
  A^q_{n0}(0) & = & \langle x^{n-1}\rangle_q \\
  \tilde{A}^q_{n0}(0) & = & \langle x^{n-1}\rangle_{\Delta q}\,.
\end{eqnarray*}
Additionally the moments of the transversity distribution, labeled by
$\delta q$ in the following, can be calculated with similar
expressions.  All the remaining details can be found
in~\cite{Dolgov:2002zm}.

All but the most trivial moments must be renormalized at some scale.
With the exception of the axial coupling, all the moments discussed
here are matched in the chiral limit to $\overline{MS}$ at
$\mu=2~\mathrm{GeV}$ using one-loop perturbation
theory~\cite{Bistrovic:2005th}.  The one exception, the axial
coupling, is renormalized non-perturbatively in the chiral limit using
the five dimensional domain wall conserved axial current.

\begin{figure}
  \begin{minipage}{210pt}
    \begin{center}
      \includegraphics[scale=0.3,clip=true]{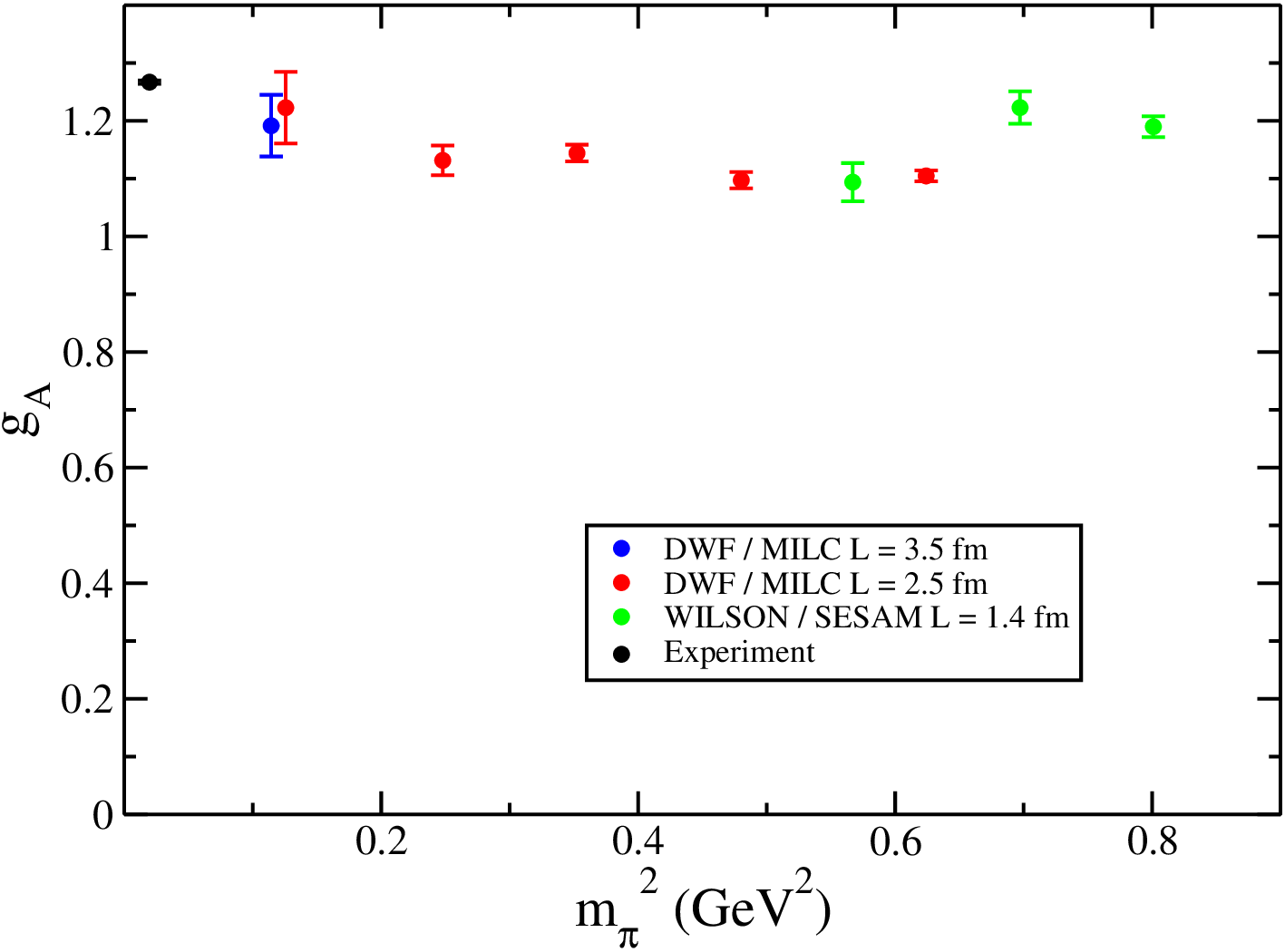}
    \end{center}
    \caption{$\langle 1\rangle_{\Delta u-\Delta d}$}
    \label{fig:g_a}
  \end{minipage}\hspace{10pt}
  \begin{minipage}{210pt}
    \begin{center}
      \includegraphics[scale=0.3,clip=true]{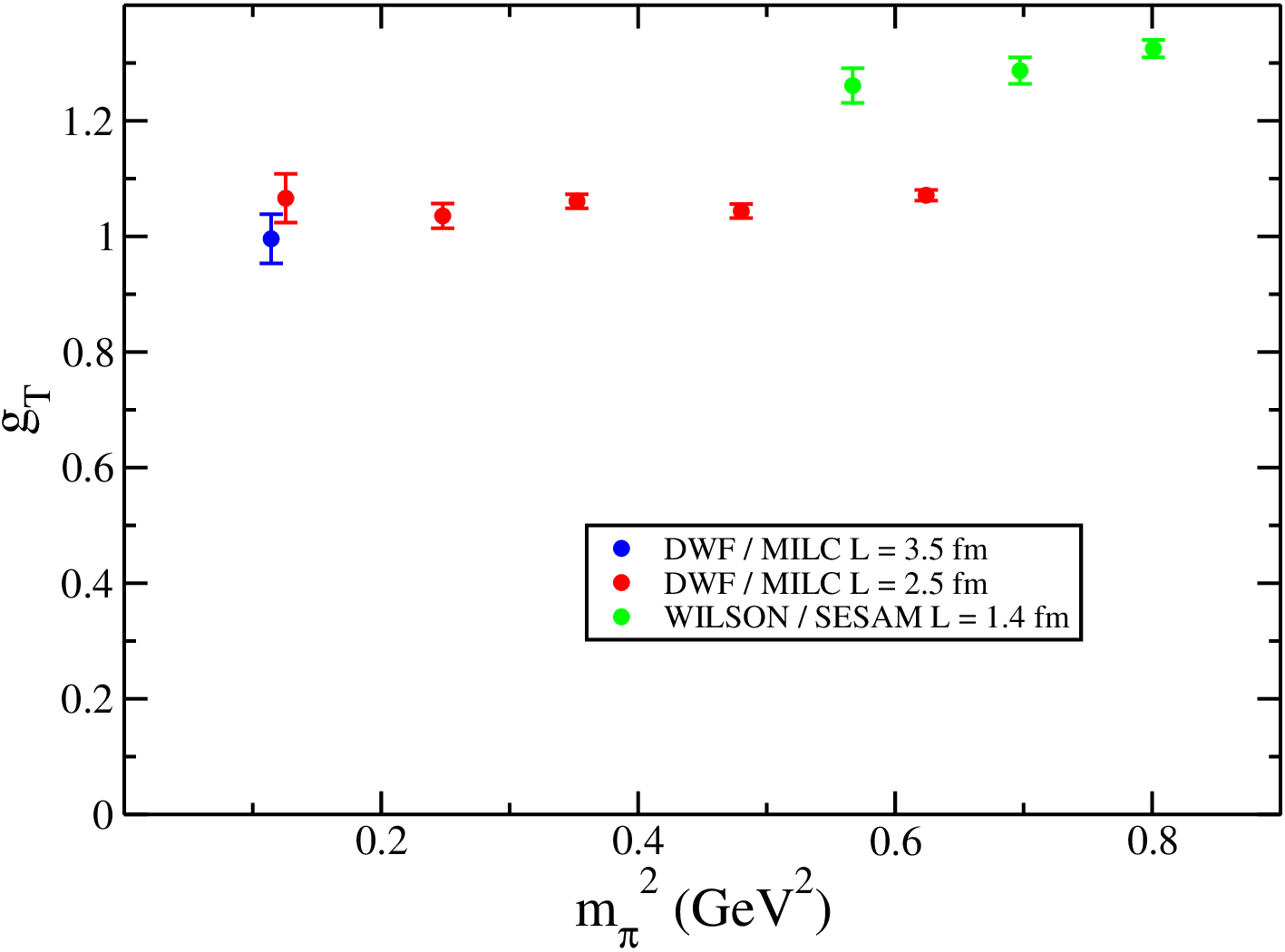}
    \end{center}
    \caption{$\langle 1\rangle_{\delta u-\delta d}$}
    \label{fig:g_t}
  \end{minipage}
\end{figure}

\begin{figure}
  \begin{minipage}{210pt}
    \begin{center}
      \includegraphics[scale=0.3,clip=true]{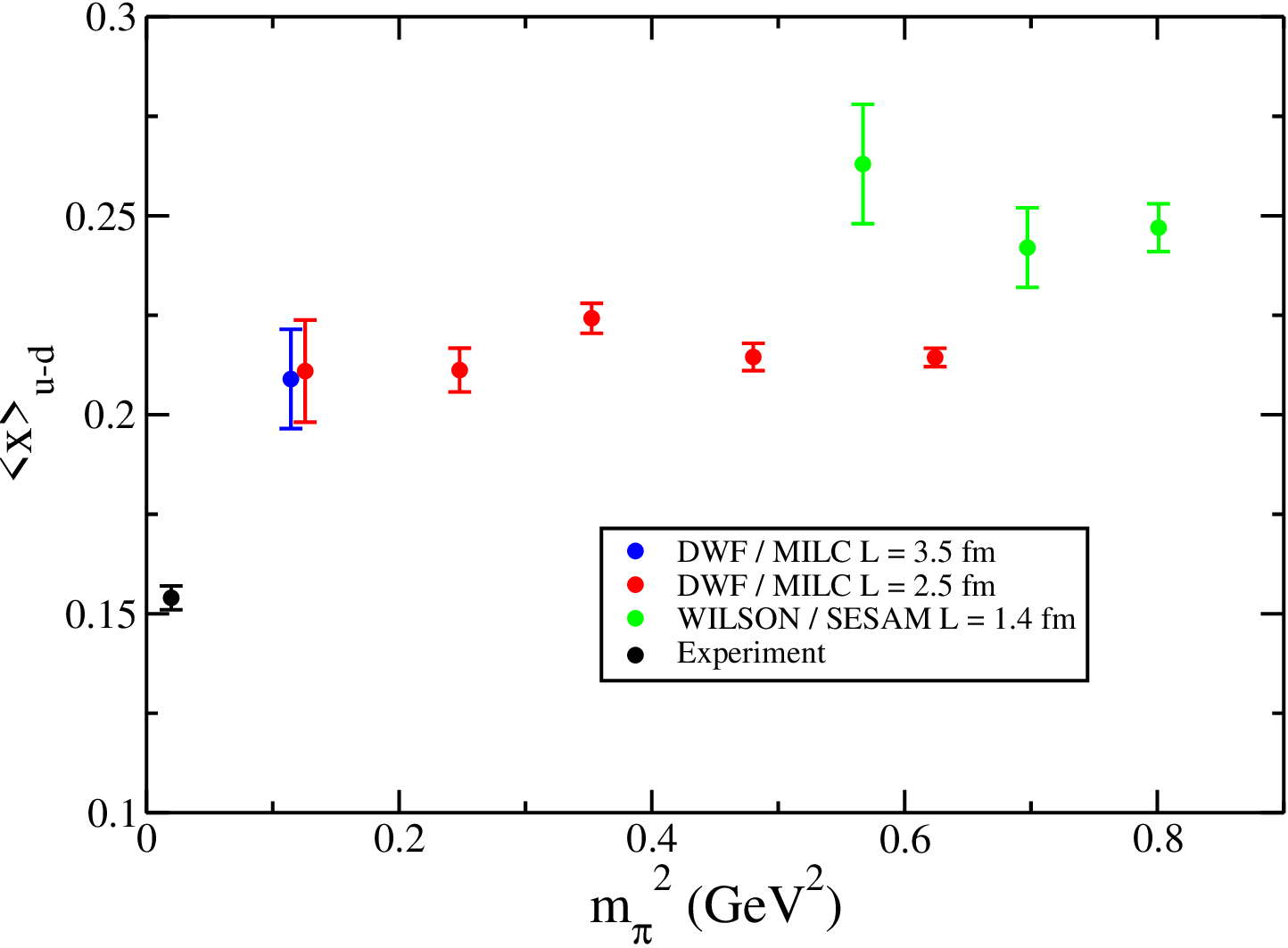}
    \end{center}
    \caption{$\langle x\rangle_{u-d}$}
    \label{fig:x}
  \end{minipage}\hspace{5pt}
  \begin{minipage}{210pt}
    \begin{center}
      \includegraphics[scale=0.3,clip=true]{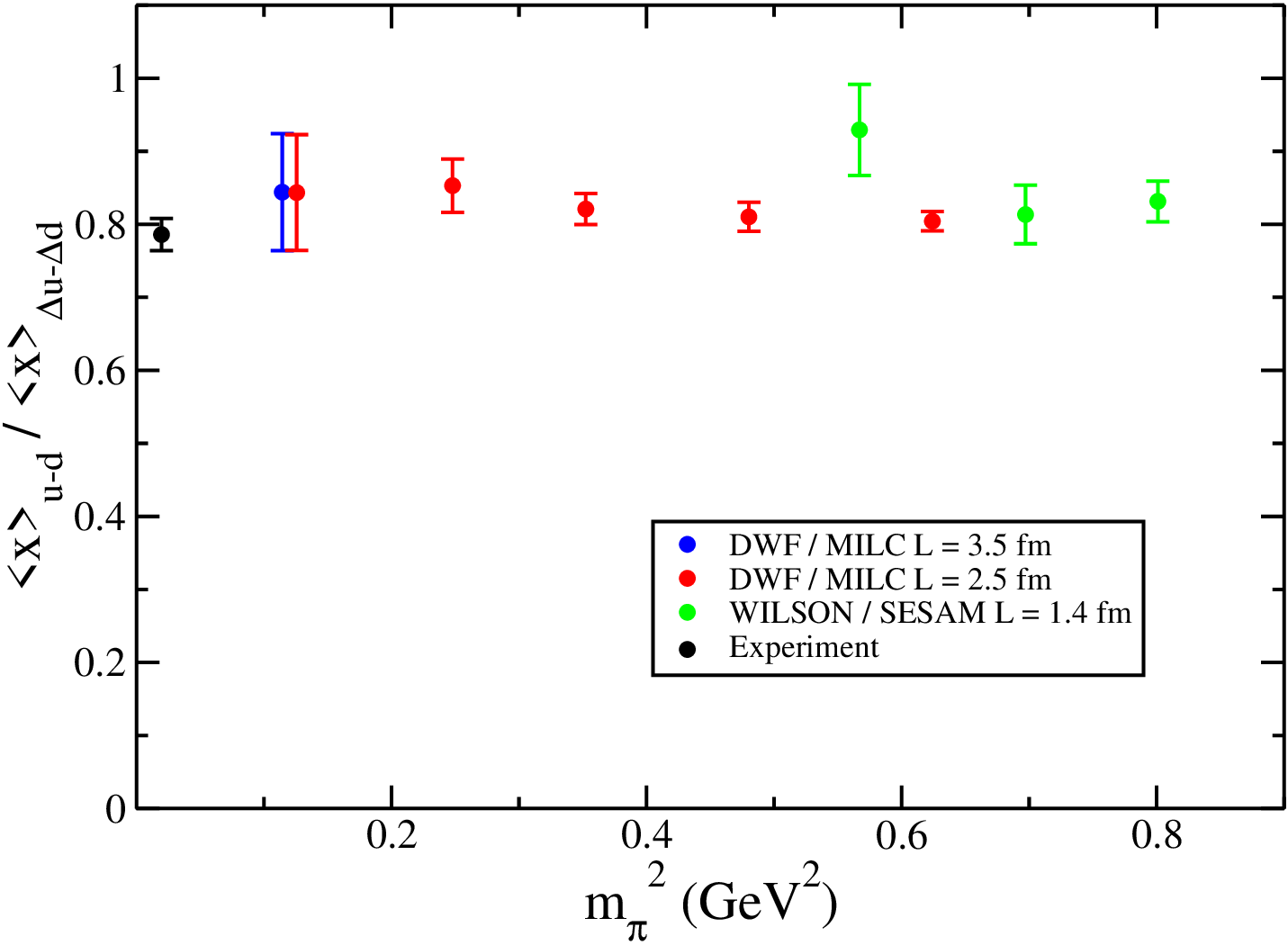}
    \end{center}
    \caption{$\langle x\rangle_{u-d}/\langle x\rangle_{\Delta u-\Delta
        d}$}
    \label{fig:x_ratio}
  \end{minipage}
\end{figure}

Figures~\ref{fig:g_a} through \ref{fig:x_ratio} show the results from
this calculation and from a previous calculation with Wilson
quarks~\cite{Dolgov:2002zm}. With one exception the points reading
from left to right denote the following: point 1 is the experimental
value, point 2 is the $3.5~\mathrm{fm}$ domain wall calculation,
points 3-6 and 8 are the $2.5~\mathrm{fm}$ domain wall calculations,
and points 7 and 9-10 are the $1.5~\mathrm{fm}$ Wilson calculations.
The one exception is that Figure~\ref{fig:g_t} lacks the experimental
value but is otherwise identical.

\paragraph{Axial and tensor couplings}
As our calculations enter the chiral regime, the axial coupling is a
particularly good benchmark observable. Physically $g_A=\langle
1\rangle_{\Delta u -\Delta d}$ represents the non-singlet contribution
of light quark spins to the nucleon spin.  Furthermore it is well
determined in neutron $\beta$ decay, and more importantly it has no
disconnected diagrams and thus is unambiguously calculable with
current lattice calculations.  Figure~\ref{fig:g_a} illustrates our
recent calculations of $g_A$ along with our group's previous
calculations at heavier quark masses~\cite{Dolgov:2002zm}.  As
mentioned earlier, the renormalization constant is calculated
non-perturbatively in the chiral limit and is determine to be
$Z_A=1.0751(11)$.  As Figure~\ref{fig:g_a} illustrates very clearly,
we are making significant progress in calculating $g_A$ at the
physical quark masses.  A tentative calculation by our group to study
the chiral extrapolation of $g_A$ as a function of the pion mass is
given in~\cite{Schroers:2005rm} and will be examined carefully in an
upcoming publication.  Additionally, the axial coupling provides a
good test of finite size effects because previous
calculations~\cite{Sasaki:2003jh} have shown that $g_A$ is
particularly sensitive to the finite size of a calculation.
Figure~\ref{fig:g_a} shows our results for volumes of
$(2.5~\mathrm{fm})^3$ and $(3.5~\mathrm{fm})^3$ at the lightest quark
mass indicating that finite size effects appear to be smaller than the
statistical accuracy of our calculations.

Our corresponding results for the non-singlet tensor charge,
$g_T=\langle 1\rangle_{\delta u - \delta d}$, are shown in
Figure~\ref{fig:g_t} at each of the same pion masses as in
Figure~\ref{fig:g_a}.  The most noteworthy observation is that the
discrepancy between the Wilson and domain wall results is greater than
for the axial charge.  Finite size effects and lattice artifacts may
account for the discrepancy, but as suggested in~\cite{QCDSF:lat05}
renormalization may also be a significant effect.  A similar mismatch
is observed for the momentum fraction as well, and we plan to study
the non-perturbative renormalization of these operators to investigate
the discrepancy.

\paragraph{Unpolarized and polarized momentum fractions}
Unlike the axial coupling, the momentum fraction remains a challenge
for contemporary lattice calculations.  Simply put, nearly all lattice
calculations to date overestimate the momentum fraction, which
represents the quark contribution to the nucleon's momentum.  Hence
lattice calculations do not yet seem to correctly account for the
sizable fraction of momentum carried by gluons in the chiral limit.
Thus it appears that the main obstacle in correctly determining the
physical value of $\langle x\rangle$ is to calculate at sufficiently
light quark masses such that chiral perturbation theory can be used to
reliably extrapolate to the physical masses~\cite{Detmold:2001jb}.
Figure~\ref{fig:x} illustrates our groups progress toward this goal.
In this figure we focus on the non-singlet contribution, $\langle
x\rangle_{u-d}$, to avoid complications arising from disconnected
diagrams.  The issues involved in correctly matching the previous
heavier Wilson quark calculations to the current domain wall
calculations were discussed earlier with regard to the tensor charge,
and the same comments apply here.  It is worthwhile to note that, even
though the lattice calculations at lighter pion masses do not yet show
any significant curvature, the lighter results do show a systematic
shift toward the experimental result.

Despite the challenges involved in $\langle x\rangle$, it was noted
in~\cite{Orginos:2005uy} that the ratio of unpolarized to
longitudinally polarized momentum fractions yields an observable which
appears to be less sensitive to the issues which still plague $\langle
x \rangle$ alone.  In particular with a chiral action the
renormalization of operators which differ solely by an insertion of
$\gamma_5$ is identical.  Hence the ratio $\langle x\rangle_{u-d} /
\langle x\rangle_{\Delta u - \Delta d}$ requires no renormalization.
Furthermore as illustrated in Figure~\ref{fig:x_ratio} the finite size
and pion mass dependence appears to strongly cancel in this ratio as
well.

\section{Moments of generalized parton distributions}
\label{sec:gpd-moments}
In the following, we only present numerical results from the lightest
working point on the $28^3\times 32$ lattice. Hence, these simulations
correspond to a pion mass of $m_\pi=352$ MeV. Furthermore, the results
presented in this section have not yet been renormalized in the
\msbar-scheme, but in the lattice regularization of our action at the
scale given in section~\ref{sec:hybrid-calculations}. They are
intended as a test of our lattice technology and not to be compared
directly to experimental data so far.

\begin{figure}
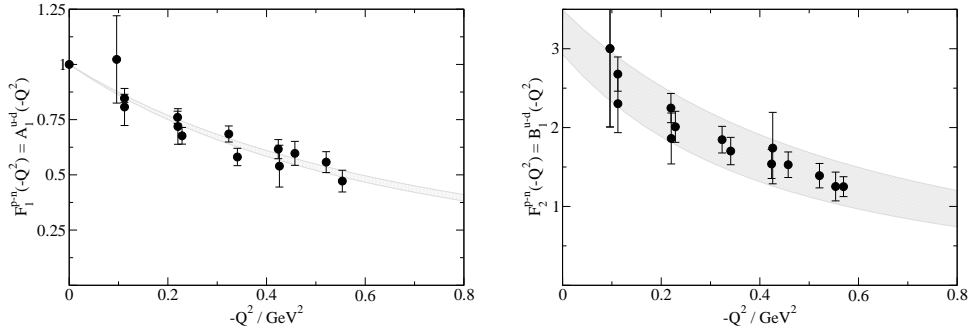

  \begin{center}
    \includegraphics[scale=0.25,clip=true]{plots/F1_u-d_28q}\hspace{15pt}
    \includegraphics[scale=0.25,clip=true]{plots/F2_u-d_28q}
  \end{center}
  \caption{The electromagnetic form factors of the nucleon at small
    values of $Q^2$.}
  \label{fig:elmagff}
\end{figure}

Our results for the electromagnetic form factors of the nucleon,
$F_1^{\mbox{\tiny p-n}}(t) = A_{10}^{\mbox{\tiny u-d}}(t)$ and
$F_2^{\mbox{\tiny p-n}}(t) = B_{10}^{\mbox{\tiny u-d}}(t)$, are shown
in figure~\ref{fig:elmagff}.

\begin{figure}
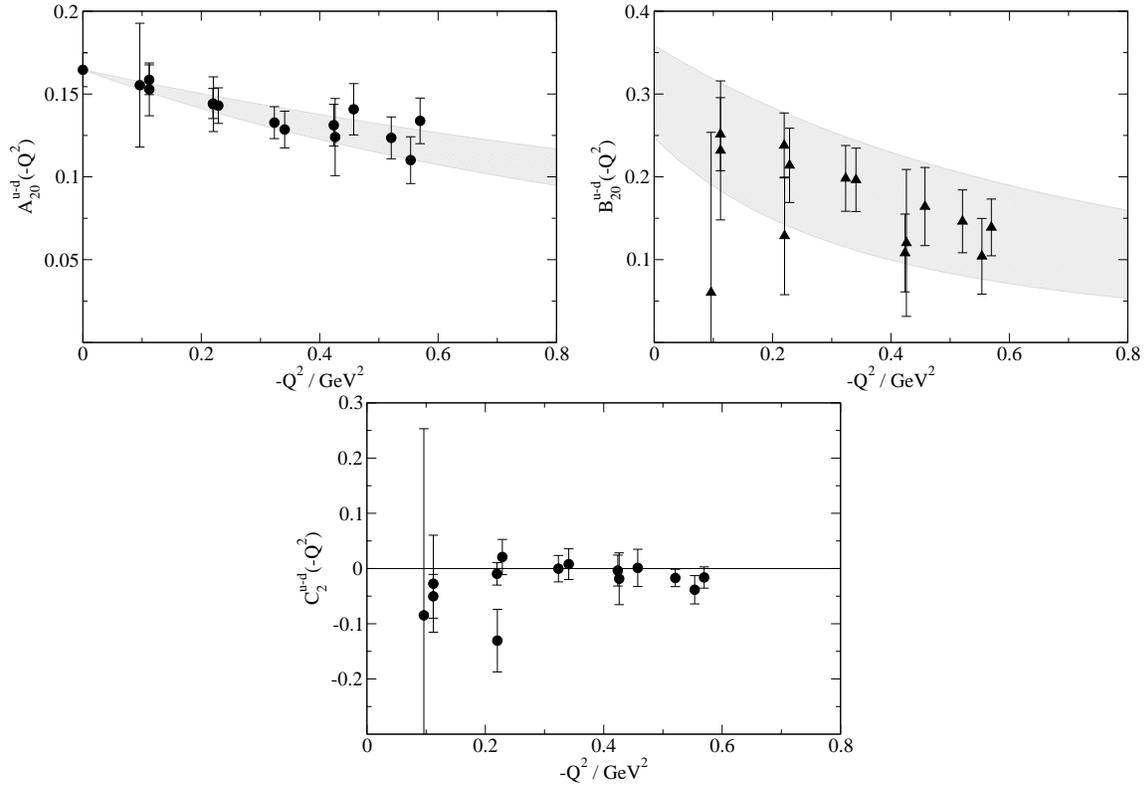

  \begin{center}
    \includegraphics[scale=0.3,clip=true]{plots/A20_u-d_28q}\hspace{5pt}
    \includegraphics[scale=0.3,clip=true]{plots/B20_u-d_28q}\\
    \includegraphics[scale=0.3,clip=true]{plots/C2_u-d_28q}
  \end{center}
  \caption{The three GFFs, $A_{20}$, $B_{20}$, and $C_2$, appearing in
    the parameterization of the nucleon energy-momentum tensor. The
    data is unrenormalized and would be corrected by about 20\% if
    renormalized in the \msbar\ scheme at a scale of $\mu_R=2$ GeV.}
  \label{fig:em-tensor}
\end{figure}

The three generalized form factors appearing in the parameterization
of the nucleon energy-momentum tensor, cf.~eq.~(\ref{eq:mom-eq}), are
shown in figure~\ref{fig:em-tensor}.

\begin{figure}
  \begin{center}
    \includegraphics[scale=0.3,clip=true]{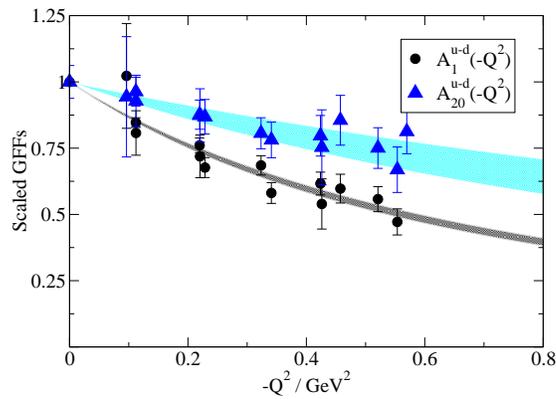}
  \end{center}
  \caption{The comparison of the scaled GFFs $A_{10}(t)$ and
    $A_{20}(t)$ signaling the non-factorization of the longitudinal
    and transverse momentum dependence.}
  \label{fig:a12-scal}
\end{figure}

An important qualitative finding of \cite{Hagler:2003is} was that the
transverse structure of the nucleon is indeed not described properly
by the assumption that the $t$-dependence factorizes from the $x$ and
$\xi$ dependence of a GPD. This ansatz has been made in several
phenomenological studies and turned out to be inaccurate. We
reproduced a key result in figure~\ref{fig:a12-scal} and find, again,
that the transverse shape of the nucleon becomes narrower as the
longitudinal momentum fraction, $x$, increases.

\section{Form factors at large momentum transfer}
\label{sec:form-factors}
\begin{figure}
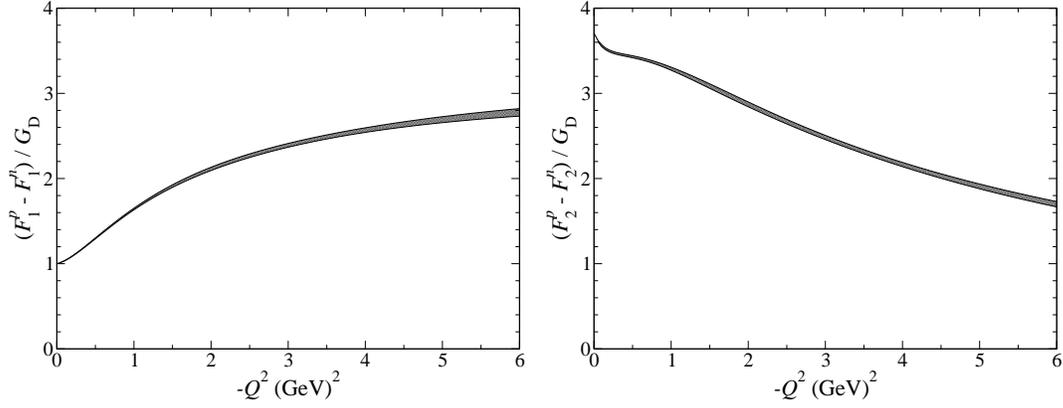

  \begin{center}
    \includegraphics[scale=0.25,clip=true]{plots/F1_p-n_expt}\hspace{5pt}
    \includegraphics[scale=0.25,clip=true]{plots/F2_p-n_expt}
  \end{center}
  \caption{\label{fig:F_p-n_expt} Experimental parameterizations of
    isovector $F_1$ (Dirac) and $F_2$ (Pauli) form factors. Dipole
    form factor $G_D$ has $\Lambda^2$=0.71 $\mathrm{GeV}^2$
    \cite{Kelly:2004hm}.}
\end{figure}

\begin{figure}
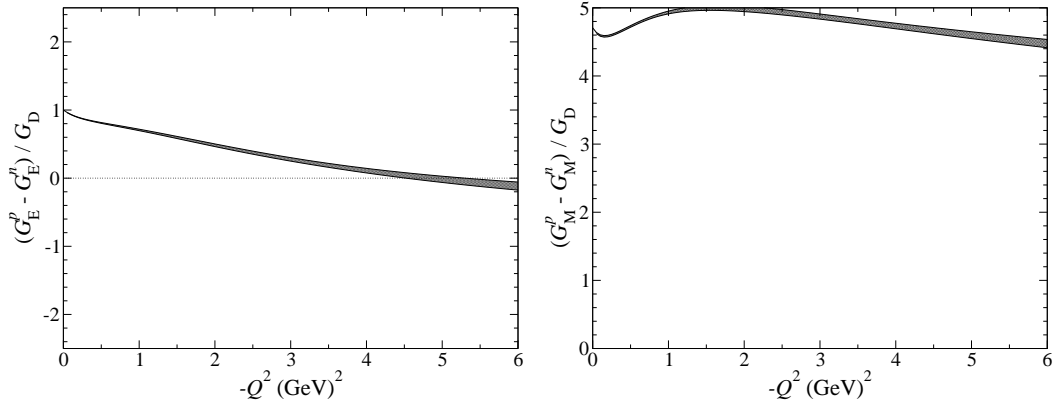

  \begin{center}
    \includegraphics[scale=0.25,clip=true]{plots/GE_p-n_expt}\hspace{5pt}
    \includegraphics[scale=0.25,clip=true]{plots/GM_p-n_expt}
  \end{center}
  \caption{\label{fig:G_p-n_expt} Experimental parameterizations of
    isovector Sachs $G_E$ (electric) and $G_M$ (magnetic) form
    factors.  Dipole form factor $G_D$ has $\Lambda^2$=0.71
    $\mathrm{GeV}^2$ \cite{Kelly:2004hm}.}
\end{figure}

\paragraph{Motivation}
The isovector part of the electromagnetic form factors of the nucleon
for momentum transfers less than 1 GeV have already been discussed in
section \ref{sec:gpd-moments} and shown in figure \ref{fig:elmagff}.
The shaded bands are a dipole fit to the lattice data and consistent
with the phenomenological notion that the form factors generally
follow the dipole form
\begin{equation}
  G_D(Q^2) \propto \left(1 - \frac{Q^2}{\Lambda^2}\right)^{-2}, \qquad
  \Lambda^2 \approx 0.71\ \mathrm{GeV}^2
\end{equation}
in this range of momentum transfers.  This can be seen directly from
the simple parameterization of the existing experimental data for
Sachs form factors by J.~J.~Kelly~\cite{Kelly:2004hm}.  In our
normalization, the Sachs form factors are determined by the Dirac and
Pauli form factors as
\begin{equation}
\label{eq:Sachs_form_factors}
G_E(Q^2) = F_1(Q^2) - \tau F_2(Q^2), \quad
G_M(Q^2) = F_1(Q^2) + F_2(Q^2), \quad
\tau = \frac{-Q^2}{4 m_N^2} \,.
\end{equation}
We can invert equation (\ref{eq:Sachs_form_factors}), form isovector
combinations and, using the parameterization \cite{Kelly:2004hm} and
its covariance matrix \cite{Kelly:2005}, produce curves derived from
experimental data for direct comparison with lattice isovector form
factors.  Figure \ref{fig:F_p-n_expt} shows the 1 $\sigma$ bands for
the Dirac and Pauli form factors.  Figure \ref{fig:G_p-n_expt} shows
similar 1 $\sigma$ bands for the Sachs electric and magnetic form
factors.

While all experimental curves show some variation with $Q^2$ relative
to the dipole form, the most intriguing feature is the apparent
zero-crossing of $G_\mathrm{E}^p - G_\mathrm{E}^n$ around $-Q^2 \approx
5\ \mathrm{GeV}^2$.  This is similar to the possible zero-crossing of
$G_\mathrm{E}^p / G_\mathrm{M}^p$ around $-Q^2 \approx 10\
\mathrm{GeV}^2$ as suggested by various extrapolations of the data
from recent polarization transfer experiments
\cite{Milbrath:1997de,Pospischil:2001pp,Gayou:2001qd,Gayou:2001qt,Punjabi:2005wq}.
It appears that by focusing on the isovector parts of form factors it
may be possible for lattice calculations to predict the vanishing of
the electric form factor at lower $Q^2$ than previously thought
possible.  Thus, the purpose of this section is to estimate what level
of statistics are required to achieve a significant measurement of a
negative electric form factor at $-Q^2 \approx 6\ \mathrm{GeV}^2$.

Another motivation for calculating form factors in the regime $-Q^2 \gg
1\ \mathrm{GeV}^2$ is to test the predictions of perturbative QCD
(pQCD) at asymptotically high $Q^2$
\cite{Brodsky:1974vy,Lepage:1980fj}.  In pQCD, the nucleon form
factors are expected to scale: $F_1 \sim Q^{-4}$, $F_2 \sim Q^{-6}$ as
$Q^2 \to \infty$.  So, it is easy to form scaling ratios which should
be roughly constant at higher momentum transfers.  Ratios like $Q^2
F_2 / F_1$ and $G_E / G_M$ should scale but have not yet been observed
to do so.  Recently, it has been noted
\cite{Brodsky:2002st,Belitsky:2002kj} that the ratio $F_2 / F_1$ has
an additional logarithmic factor: $Q^2 F_2 / F_1 \sim
\log^2(Q^2/\Lambda^2)$ where $\Lambda \approx 300\ \mathrm{MeV}$ seems
to roughly restore scaling of experimental data in the range
$-Q^2 = 2\ \mbox{--}\ 6\ \mathrm{GeV}^2$.

A word of caution is in order with regards to apparent scaling
of the experimental data having its origins in pQCD scaling
in the asymptotic regime.  In a recent lattice calculation
of the pion electromagnetic form factor $F_\pi(Q^2)$ \cite{Bonnet:2004fr}
in the same hybrid scheme as described in section \ref{sec:hybrid-calculations},
scaling was observed for $F_\pi(Q^2)$ as predicted by pQCD.  In the case
of the pion, it is also possible to compute the asymptotic normalization
and the data did not yet agree with the pQCD prediction.  Thus, scaling alone
is not sufficient to establish the reliability of pQCD calculations
of nucleon form factors.

\begin{figure}
  \begin{center}
    \includegraphics[scale=0.25,clip=true]{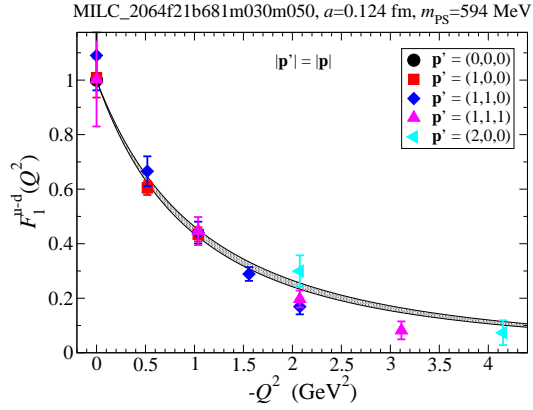}
  \end{center}
  \caption{\label{fig:Breit} Isovector Dirac form factor in the Breit
    frame $( \left| \vec{p}^\prime \right| = \left| \vec{p} \right|)$
    for several sink momenta $\vec{p}^\prime$ \textit{vs.}\ a dipole
    fit of the data.}
\end{figure}

\paragraph{Exploring higher $Q^2$ on the lattice}
To estimate the computational costs of computing nucleon electromagnetic
form factors to $-Q^2 \sim 6\ \mathrm{GeV}^2$, we computed sequential
propagators on 282 configurations separated by 12 MILC HMD trajectories,
at $m_\mathrm{PS} = 594(2)\ \mathrm{MeV}$,
half of those listed in table \ref{tab:hybrid-stat}, for higher sink
momenta $\vec{p}^\prime$ = (1,1,0), (1,1,1) and (2,0,0).  For this study,
we focused on computing in the Breit frame
$( \vec{p}^\prime = - \vec{p} )$ as past experience
has shown that computed form factors have smaller statistical variance than
other momentum combinations at the same $Q^2$.  Our results
for $F_1^{u-d}(Q^2)$ are shown in figure \ref{fig:Breit} and are consistent
with dipole scaling over the entire range of $Q^2$.

\begin{figure}
  \begin{center}
    \includegraphics[scale=0.25,clip=true]{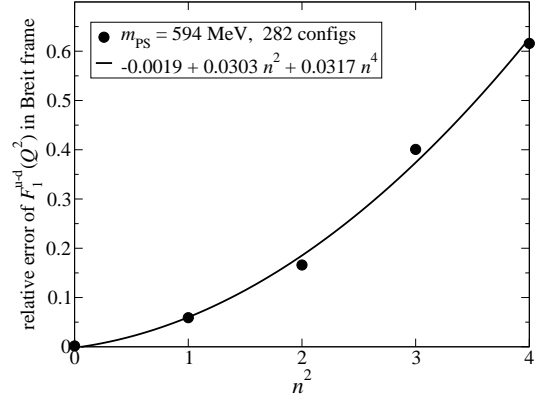}
  \end{center}
  \caption{\label{fig:rel_vs_mom2} Relative error of the isovector
    Dirac form factor \textit{vs.}\ sink momenta $\vec{p}^\prime = 2
    \pi \vec{n} / L$, where $n^2 = \vec{n} \cdot \vec{n}$, at fixed
    $m_\mathrm{PS} = 594(2)\ \mathrm{MeV}$.}
\end{figure}

In figure \ref{fig:rel_vs_mom2}, we plot the relative error
of $F_1^{u-d}(Q^2)$ in the Breit frame.  We fit the points
to several functional forms with three or less free parameters
in order to extrapolate our results to higher sink momenta.
The polynomial curve shown was the best representation
of the data we found.  This form is motivated by the following picture.
$F_1$ decreases as $Q^{-4} \propto n^{-4}$ in the Breit frame.
If the statistical noise in our matrix element construction is independent
of the magnitude of the sink momenta, then the relative error should increase
as $n^4$ which is consistent with our calculation.

Because we would like to establish whether a zero-crossing
occurs around $-Q^2 = 5\ \mathrm{GeV}^2$ for the isovector electric form factor,
we would like to set a goal of achieving 30\% relative errors up
to 6 $\mathrm{GeV}^2$.  For $\vec{p}^\prime$ = (2,0,0), $n^2$ = 4
and $-Q^2 = 4.15\ \mathrm{GeV}^2$ the relative error is 62\%.
Reducing the relative error to 30\% at this $Q^2$ should therefore require
four times the number of configurations, two times more than our current dataset.

To reach 6 $\mathrm{GeV}^2$ requires a sink momenta of $\vec{p}^\prime$ = (2,2,0),
$n^2$ = 8 and our extrapolation predicts we should find a relative error
of 227\% on our set of 282 configurations.  To reduce this relative error
to 30\% should require 57 times as many configurations, on the order
of 200,000 MILC HMD trajectories!  Since it is unlikely that this ensemble will
be extended by so many trajectories, achieving our goals will require some other
technique to reduce the variance of matrix elements at higher $Q^2$.

\begin{figure}
  \begin{center}
    \includegraphics[scale=0.25,clip=true]{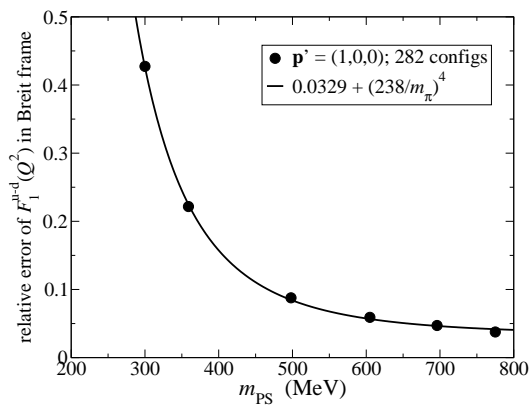}
  \end{center}
  \caption{\label{fig:rel_vs_mpi} Relative error of the isovector
    Dirac form factor \textit{vs.}\ pion mass at fixed sink momentum.
  }
\end{figure}

Another important question is how does the relative error of the form factors
depend on the dynamical pion mass at fixed $Q^2$.  In figure \ref{fig:rel_vs_mpi}
we show the relative error for $F_1^{u-d}(Q^2)$ as a function of the pion mass
in the Breit frame at fixed sink momenta $\vec{p}^\prime$ = (1,0,0).  Note that
the box size is held fixed in lattice units ($L$=20) and thus $m_\mathrm{PS} L$
varies but always $m_\mathrm{PS} L > 4$.  As before, we attempted to fit
various functional forms to the data and the form which best fit the data
is plotted.  In this case, we were unable to find any other functional form with
three or fewer parameters which could fit the data with any comparable accuracy.
We do not currently have a good physical motivation for this functional form
except to note that the scale at which the divergence sets in is
approximately $250\ \mathrm{MeV}$, comparable with the scale at which other chiral
extrapolations of hadronic matrix elements typically show substantial curvature.
This figure clearly reiterates the need to identify some other technique
for reducing the variance of hadronic matrix elements at higher $Q^2$.

\section{Summary and outlook}
\label{sec:summary-outlook}
We have demonstrated selected results for the structure of the nucleon
at light quark masses using our hybrid approach. We have focused on a
specific sample at one working point of our simulations to demonstrate
the qualitative behavior of the generalized form factors. The results
are in qualitative agreement with previous findings and we did not
encounter major problems during our runs. We then focused on a direct
comparison of several moments of the forward limit of parton
distributions to experiment. These results included our entire data
set and the direct comparison to experimental data has given
quantitative agreement in several cases once appropriate chiral
extrapolations have been considered. Finally, we have explored the
possibility of exploring the regime of larger virtual momentum
transfers in the case of the nucleon form factors.

The encouraging results make us confident that the hybrid approach is
indeed promising to bridge the gap between current lattice data and
the regime where chiral perturbation theory is applicable. These
results are therefore of great importance to the qualitative and
quantitative understanding of hadronic matrix elements.

Before the advent of light dynamical overlap and domain-wall
calculations we will provide a complete analysis of selected
quantities like the nucleon axial coupling \cite{LHPC:2005pr}, light
hadron spectroscopy \cite{LHPC:2005ps}, and all accessible moments of
generalized parton distributions \cite{LHPC:2005pt}.

\acknowledgments
This work was supported in part by the DOE Office of Nuclear Physics
under contracts DE-FC02-94ER40818, DE-FG02-92ER40676, and DE-
AC05-84ER40150. It was also supported in part by the EU Integrated
Infrastructure Initiative Hadron Physics (I3HP) under contract
RII3-CT-2004-506078 and by the DFG under contract FOR 465
(Forschergruppe Gitter-Hadronen-Ph\"anomenologie). Computations were
performed on clusters at Jefferson Laboratory and at ORNL using time
awarded under the SciDAC initiative. We are indebted to members of the
MILC and SESAM collaborations for the dynamical quark configurations
which made our full QCD calculations possible. WS wishes to thank
James Zanotti for carefully reading this manuscript and Thomas Streuer
for discussions.

\appendix

%
%


\begin{thebibliography}{999}
\bibitem{Muller:1998fv} D.~M{\"u}ller, D.~Robaschik, B.~Geyer,
  F.~M.~Dittes and J.~Horejsi, Fortsch.\ Phys.\  {\bf 42}, 101
  (1994). \\
  X.~D.~Ji, Phys.\ Rev.\ Lett.\  {\bf 78}, 610 (1997). \\
  A.~V.~Radyushkin, Phys.\ Rev.\ D {\bf 56}, 5524 (1997).
\bibitem{Diehl:2003ny} M.~Diehl, Phys.\ Rept.\ {\bf 388}, 41 (2003). \\
  A.~V.~Radyushkin, Phys.\ Rev.\ D {\bf 56}, 5524 (1997).
\bibitem{Mathur:1999uf} N.~Mathur, S.~J.~Dong, K.~F.~Liu,
  L.~Mankiewicz and N.~C.~Mukhopadhyay, Phys.\ Rev.\ D {\bf 62},
  114504 (2000).
\bibitem{Gockeler:2003jf} M.~G{\"o}ckeler, R.~Horsley, D.~Pleiter,
  P.~E.~L.~Rakow, A.~Sch{\"a}fer, G.~Schierholz and W.~Schroers [QCDSF
  Collaboration], Phys.\ Rev.\ Lett.\ {\bf 92}, 042002 (2004).
\bibitem{Hagler:2003jd} P.~H{\"a}gler, J.~Negele, D.~B.~Renner,
  W.~Schroers, T.~Lippert and K.~Schilling [LHPC Collaboration],
  Phys.\ Rev.\ D {\bf 68}, 034505 (2003).
\bibitem{Diehl:2004cx} M.~Diehl, T.~Feldmann, R.~Jakob and P.~Kroll,
  Eur.\ Phys.\ J.\ C {\bf 39}, 1 (2005).
\bibitem{Hagler:2003is} P.~H{\"a}gler, J.~W.~Negele, D.~B.~Renner,
  W.~Schroers, T.~Lippert and K.~Schilling [LHPC Collaboration],
  Phys.\ Rev.\ Lett.\ {\bf 93}, 112001 (2004). \\
  M.~G{\"o}ckeler {\it et al.}  [QCDSF Collaboration], Nucl.\ Phys.\
  Proc.\ Suppl.\ {\bf 128}, 203 (2004). \\
  M.~G{\"o}ckeler {\it et al.}  [QCDSF Collaboration], Nucl.\ Phys.\
  Proc.\ Suppl.\ {\bf 135}, 156 (2004). \\
  J.~W.~Negele {\it et al.}  [LHPC Collaboration], Nucl.\ Phys.\
  Proc.\ Suppl.\ {\bf 129}, 910 (2004). \\
  B.~Bistrovic {\it et al.}  [LHPC Collaboration],
  arXiv:hep-lat/0509101.
\bibitem{Schroers:2003mf} W.~Schroers {\it et al.}  [LHPC
  collaboration], Nucl.\ Phys.\ Proc.\ Suppl.\ {\bf 129}, 907
  (2004). \\
  P.~H{\"a}gler, J.~W.~Negele, D.~B.~Renner, W.~Schroers, T.~Lippert
  and K.~Schilling [LHPC Collaboration], Eur.\ Phys.\ J.\ A {\bf
    24S1}, 29 (2005).
\bibitem{Gockeler:2005aw} M.~G{\"o}ckeler {\it et al.}  [QCDSF
  Collaboration], arXiv:hep-lat/0501029. \\
  M.~G{\"o}ckeler {\it et al.}  [QCDSF Collaboration],
  arXiv:hep-lat/0507001. \\
  A.~Sch{\"a}fer {\it et al.} [QCDSF Collaboration], these
  proceedings.
\bibitem{Renner:2004ck} D.~B.~Renner {\it et al.}  [LHP
  Collaboration], Nucl.\ Phys.\ Proc.\ Suppl.\ {\bf 140}, 255 (2005).
\bibitem{Brommel:2005la} D.~Br{\"o}mmel {\it et al.},
  arXiv:hep-lat/0509133. These proceedings.
\bibitem{Dolgov:2002zm} D.~Dolgov {\it et al.}  [LHPC Collaboration],
  Phys.\ Rev.\ D {\bf 66}, 034506 (2002).
\bibitem{Bistrovic:2005th} B.~Bistrovic {\it et al.}, (2005), in
  preparation.
\bibitem{Schroers:2005rm}   D.~B.~Renner, arXiv:hep-lat/0508008. \\
  D.~B.~Renner, J.\ Phys.\ Conf.\ Ser.\ {\bf 9}, 264 (2005). \\
  W.~Schroers, Nucl.\ Phys.\ A {\bf 755}, 333 (2005).
\bibitem{Sasaki:2003jh} S.~Sasaki {\it et al.}  [RBCK Collaboration],
  Phys.\ Rev.\ D {\bf 68}, 054509 (2003).
\bibitem{QCDSF:lat05} T.~Streuer {\it et al.} [QCDSF Collaboration],
  these proceedings.
\bibitem{Detmold:2001jb} W.~Detmold, W.~Melnitchouk, J.~W.~Negele,
  D.~B.~Renner and A.~W.~Thomas, Phys.\ Rev.\ Lett.\ {\bf 87}, 172001
  (2001).
\bibitem{Orginos:2005uy} K.~Orginos, T.~Blum and S.~Ohta,
  arXiv:hep-lat/0505024.
\bibitem{Kelly:2004hm} J.~J.~Kelly, Phys.\ Rev.\ C {\bf 70}, 068202
  (2004).
\bibitem{Kelly:2005} J.~J.~Kelly, private communication.
\bibitem{Milbrath:1997de} B.~D.~Milbrath {\it et al.}  [Bates FPP
  collaboration], Phys.\ Rev.\ Lett.\ {\bf 80}, 452 (1998)
  [Erratum-ibid.\ {\bf 82}, 2221 (1999)].
\bibitem{Pospischil:2001pp} T.~Pospischil {\it et al.}  [A1
  Collaboration], Eur.\ Phys.\ J.\ A {\bf 12}, 125 (2001).
\bibitem{Gayou:2001qd} O.~Gayou {\it et al.}  [Jefferson Lab Hall A
  Collaboration], Phys.\ Rev.\ Lett.\ {\bf 88}, 092301 (2002).
\bibitem{Gayou:2001qt} O.~Gayou {\it et al.}, Phys.\ Rev.\ C {\bf 64},
  038202 (2001).
\bibitem{Punjabi:2005wq} V.~Punjabi {\it et al.}, Phys.\ Rev.\ C {\bf
    71}, 055202 (2005) [Erratum-ibid.\ C {\bf 71}, 069902 (2005)].
\bibitem{Brodsky:1974vy} S.~J.~Brodsky and G.~R.~Farrar, Phys.\ Rev.\
  D {\bf 11}, 1309 (1975).
\bibitem{Lepage:1980fj} G.~P.~Lepage and S.~J.~Brodsky, Phys.\ Rev.\ D
  {\bf 22}, 2157 (1980).
\bibitem{Brodsky:2002st} S.~J.~Brodsky, arXiv:hep-ph/0208158.
\bibitem{Belitsky:2002kj} A.~V.~Belitsky, X.~d.~Ji and F.~Yuan, Phys.\
  Rev.\ Lett.\ {\bf 91}, 092003 (2003).
\bibitem{Bonnet:2004fr}
  F.~D.~R.~Bonnet, R.~G.~Edwards, G.~T.~Fleming, R.~Lewis and D.~G.~Richards
  [Lattice Hadron Physics Collaboration],
  Phys.\ Rev.\ D {\bf 72}, 054506 (2005).
\bibitem{LHPC:2005pr} [LHPC Collaboration], in preparation.
\bibitem{LHPC:2005ps} [LHPC Collaboration], in preparation.
\bibitem{LHPC:2005pt} [LHPC Collaboration], in preparation.
\end{thebibliography}
\end{document}